# Numerical analysis of the strain distribution in skin domes formed upon the application of hypobaric pressure

Daniel Sebastia-Saez[1], Faiza Benaouda[2], Charlie Lim[2], Guoping Lian[1,3], Stuart Jones[2], Tao Chen[1], Liang Cui[4,*]

[1]Department of Chemical and Process Engineering, University of Surrey, United Kingdom

[2]Institute for Pharmaceutical Science, King's College London, United Kingdom

[3]Unilever R&D Colworth, United Kingdom

[4]Department of Civil and Environmental Engineering, University of Surrey, United Kingdom

*Corresponding author: Dr Liang Cui (E-mail: l.cui@surrey.ac.uk)

## Abstract

Suction cups are widely used such as in measurement of mechanical properties of skin in vivo, in drug delivery devices, or in acupuncture treatment. Understanding mechanical response of skin under hypobaric pressure are of great importance for users of suction cups. The aims of this work are to assess the capability of linear elasticity (Young's modulus) or hyperelasticity in predicting hypobaric pressure induced 3D stretching of the skin. Using experiments and computational Finite Element Method modelling, this work demonstrated that although it was possible to predict the suction dome apex height using both linear elasticity and hyperelasticity for the typical range of hypobaric pressure in medical applications (up to -10 psi), linear elasticity theory showed limitations when predicting the strain distribution across the suction dome. The reason is that the stretch ratio reaches values exceeding the initial linear elastic stage of the stress-strain characteristic curve for skin. As a result, the linear elasticity theory overpredicts the stretch along the rim of domes where there is stress concentration. In addition, the modelling showed that the skin was compressed consistently along the thickness direction, leading to reduced thickness. Using hyperelasticity modelling to predict the 3D strain distribution paves the way to accurately design safe commercial products that interface with the skin.

# LIST OF SYMBOLS

Latin characters

| | | |
|---|---|---|
| $A$ | Constant Timoshenko model | [-] |
| $a_d$ | Radius of the dome | [mm] |
| $a_1, a_2, a_3$ | Principal stretch ratios | [-] |
| $B$ | Constant Timoshenko model | [-] |
| $C$ | Fourth-order stiffness tensor | [N·m$^{-1}$] |
| $e$ | Skin thickness | [mm] |
| $E$ | Suction Young's modulus | [MPa] |
| $E_{max}$ | Maximum Young's modulus | [MPa] |
| $E_0$ | Initial Young's modulus | [MPa] |
| $\boldsymbol{f}$ | Body force per unit volume | [] |
| $\boldsymbol{F}$ | Deformation gradient tensor | [m$^{-1}$] |
| $p$ | Pressure | [Pa] |
| $r$ | Radial coordinate | [mm] |
| $R$ | Coefficient of determination | [-] |
| $\boldsymbol{u}$ | Displacement field | [mm] |
| $u_a$ | Dome apex height | [mm] |
| $W$ | Strain energy density function | [J·m$^{-3}$] |
| $z$ | Axial coordinate | [mm] |

Greek characters

| | | |
|---|---|---|
| $\alpha$ | Ogden model constant | [-] |
| $\bar{\bar{\epsilon}}$ | Strain tensor | [-] |
| $\epsilon_{rr}, \epsilon_{\phi\phi}, \epsilon_{zz}$ | Principal components strain tensor | [-] |
| $\phi$ | Angular coordinate | [rad] |

| $\mu$ | Ogden model constant | [Pa] |
| $\boldsymbol{\sigma_c}$ | Cauchy stress tensor | [Pa] |
| $\sigma$ | Stress | [Pa] |
| $\lambda$ | Stretch ratio | [-] |
| $\upsilon$ | Poisson ratio | [-] |

## 1. Introduction

Suction cups with hypobaric pressure are widely adopted in various devices to interact with skin for various medical /therapeutic/cosmetic purposes, such as Cutometer® for skin elasticity analysis, drug delivery devices, and suction cups in acupuncture treatment or spa treatment (Dobrev, 2017; Xiaoxuan Zhang et al., 2020). Suction domes with various extent of skin deformation are formed upon application of hypobaric pressure. Amongst all studies, the relationship between the pressure and skin deformation is crucial for safely designing these devices without causing damage to skin. The key mechanical property linking these two is skin elasticity.

Testing the elasticity of skin is crucial in many fields. For example, it can be used to monitor skin disease such as systemic sclerosis, burn treatment or scar rehabilitation (Elrod et al., 2019; Kumánovics et al., 2017; Xiaoming Zhang et al., 2020). Other applications include the development of real-time haptic devices to train plastic surgeons (Lapeer et al., 2010) or the development of surgical methods to avoid leaving scars after surgery (Laiacona et al., 2019). Also, the effect of different cosmetics on the skin and tactile perception can be evaluated by measuring the elasticity of skin (Park et al., 2019; Sergachev et al., 2019).

There are several approaches to test skin elasticity, but deformation testing is most commonly employed to assess the mechanical properties of skin in-vivo, including extension (Delalleau et al., 2008), torsion (Sanders, 1973), suction (Hendriks et al., 2003) and indentation testing (Griffin et al., 2016; Groves et al., 2013; Joodaki and Panzer, 2018; Lim et al., 2011). These mechanical tests can be coupled with imaging techniques and numerical modelling using Finite Element Method (FEM) to map live stress distributions on skin in-vivo. For example, a three-camera optical motion analysis system has been used to study the viscoelastic properties of skin (Mahmud et al., 2010). Digital image correlation and a single-term Ogden hyperelastic FEM model have also been combined to study the deformation of skin in-vivo. However, in-vivo testing does not provide clear boundary conditions. Tensile tests on excised samples of skin on the other hand allow the observation of greater strains.

In all these test typologies, substantial disparities have been observed in the ranges of Young's modulus observed (Pissarenko and Meyers, 2020). Kalra et al. (A and A, 2016) for instance, found that the Young's modulus obtained upon suction on human skin ranged from 0.025 MPa to 0.26 MPa. Human skin tensile tests also report wide ranges of values for the maximum slope of the stress-strain curve between 4.02 MPa and 140 MPa, depending on the strain rate range considered, i.e. quasi-static or dynamic deformation. Kalra et al. (A and A, 2016) pointed out that differences between suction and tensile Young's moduli are caused by the unequal distribution of stresses that follow both tensile and suction tests. Annaidh et al. (Ní Annaidh et al., 2012), however, suggested that suction Young's moduli are comparable to the initial slope found during tensile tests because the range of strains observed upon suction coincides with that of the early stage of a tensile test.

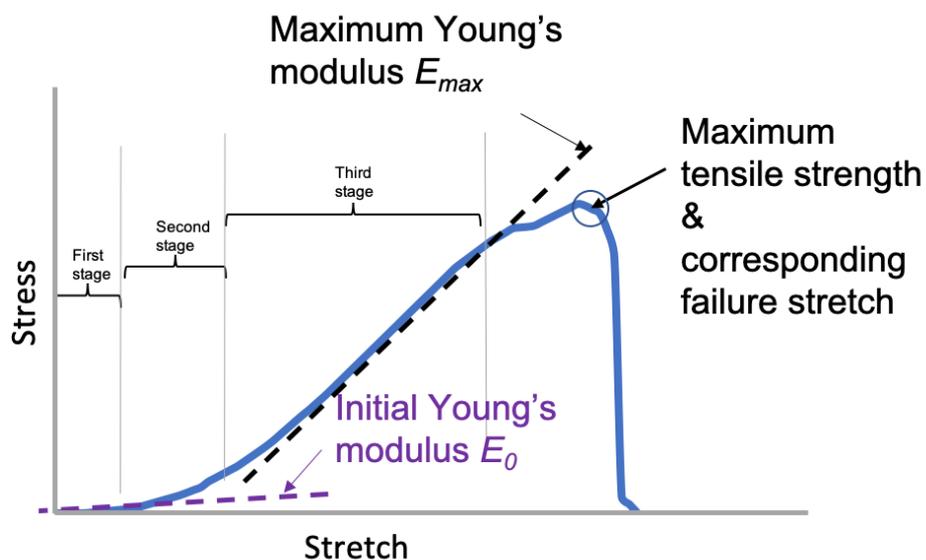

*Figure 1 Schematic illustration of a standard stress-strain curve for skin and the terminology used in this work.*

One reason for the large disparities in the skin elasticity measurements could be the use of the Young's modulus as the comparative index. The Young's modulus cannot be calculated accurately when the stress-strain response of the material is not linear (Pawlaczyk et al., 2013) and the mechanical behaviour of skin is known to be non-linear. The skin could be more accurately described, as is the case for many other soft biological tissues, by means of a non-linear elastic (hyperelastic) model. Several options, such as the Ogden, Yeoh, neo-Hookean et cetera, have given rise to good stress-strain prediction in non-linear materials, but they have not, as yet, been readily adopted for the analysis of skin elasticity (Lakhani et al., 2020). This may be because the skin has a unique structure that results in a three stage stress-strain profile that arises from the re-arrangement of fibres made of oxylatan and elaunin (in the papillary dermis) and fibrillin-core elastin fibres (in the reticular dermis) as the applied stress increases (A and A, 2016; Aziz et al., 2016; Markenscoff and Yannas, 1979; Shergold et al., 2006; Ueda et al., 2019), which adds complexity to the modelling (Figure 1).

The three stage stress-strain relationship for skin has been described by Benítez and Montáns (Benítez and Montáns, 2017). In the first stage, which can be approximated as linear, large strains are obtained upon the application of small stresses. The first stage occurs because the protein fibres are loose within the dermic matrix, allowing a softer response. The stress-strain curve slope in the first stage gives good approximation of the Young's moduli measured upon suction conditions (Edwards and Marks, 1995; Ní Annaidh et al., 2012). The second stage is non-linear and marks the transition between the first 'softer' stage mentioned above and the third 'stiffer' stage. The third stage is also characterised by a linear response, although here, a given applied stress gives rise to a smaller strain than in the first stage. This is caused by the protein fibres being aligned with the direction of stress, giving way to a stiffer response along the direction of fibre alignment than that perpendicular to the alignment direction (A and A, 2016; Holzapfel, 2000; Oxlund et al., 1988; Silver et al., 2001). As previously mentioned, there is confusion in the literature regarding which stage of the skin stretching curve should be modelled, with reports of Young's moduli being classified as representing skin elasticity when measured in both stage 1 and stage 3 of the stress-strain curve. Hyperelastic theory could be used to predict skin elasticity through more than a single stage of the stress-strain response and thus provide a more reliable prediction of 3D skin stretching compared to linear theory modelling, i.e., Young's Modulus. This would expand the capability of ex-vivo testing to predict in-vivo skin elasticity, but the merits of hyperelastic theory compared to linear theory at stresses relevant to commercial product development need further investigation.

The aim of this work was to compare linear elasticity (Young's modulus) and hyperelasticity modelling of stress-strain data from tensile tests to predict hypobaric pressure induced 3D stretching of the skin. Numerical modelling using FEM was selected because it provides a reliable and non-invasive method to obtain comprehensive information on strain distributions. The research landscape described above suggests that, to accurately predict the strain distribution in skin, the stress-strain behaviour should be characterised for the entire strain range. To achieve this, experimental tensile tests will be carried out on porcine skin samples to characterise the stress-strain characteristics for model calibration. Both linear elastic and non-linear elastic models were used to simulate the experimental suction tests to assess their validity. 3D spatial strain distributions in the skin dome in both cases were examined and its potential impact on skin damage and applications of medical devices was discussed.

## 2. Methodology and work programme

The methodology and work programme of the present work are schematised in Figure *2* and details on the tensile and suction experiments and the FEM simulations are given as follows.

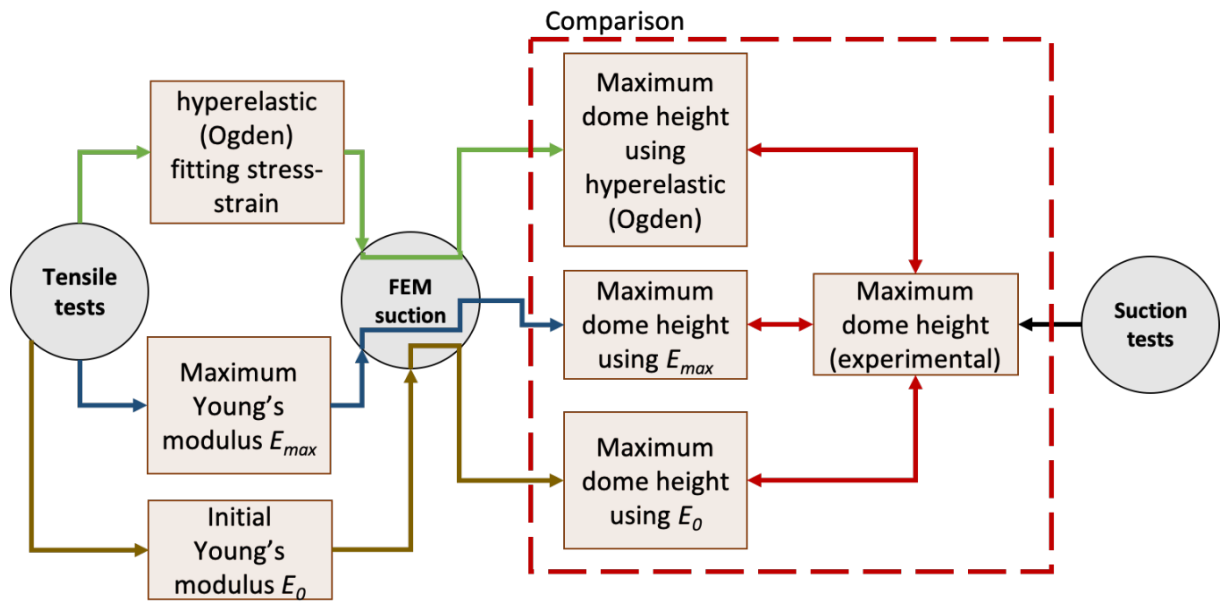

*Figure 2 Flow chart of the present work. Green route: the stress-strain relationship from the experimental tensile tests is fitted to a non-linear hyperelastic (Ogden) model and fed into the FEM simulation of the suction test; Brown route: the initial Young's modulus $E_0$ (initial slope of the tensile stress-strain curve) is fed into the FEM simulation of the suction test; Blue route: the maximum Young's modulus $E_{max}$ (maximum slope of the tensile stress-strain curve) is fed into the FEM simulation of the suction test; Red route: The dome apex heights from the three routes above are compared with those from experimental suction tests to validate the elastic parameters; 3D strain distribution in domes in FEM simulations are examined.*

## 2.1. Tensile tests

Tensile tests on "dog-bone" shaped porcine skin specimens were carried out with the dimensions and specifications described in the ASTM D-412 standard ("ASTM D412-15 Standard Test Methods for Vulcanized Rubber and Thermoplastic Elastomers," n.d.) (Figure 3a). Pig skin was selected as it is the preferred surrogate for human skin (Edwards and Marks, 1995; Ranamukhaarachchi et al., 2016; Shergold et al., 2006). The samples were prepared so that the stresses could be applied in the directions parallel or perpendicular to the Langer's lines (Figure 3c). The Langer's lines mark the alignment of the collagen fibres within the dermic matrix (Gibson, 1978; Murphy et al., 2017). A schematic of the Langer's lines can be seen in  Figure 3b and Figure 3c. The speed of the test was 50 mm/min (quasi-static) in a pull-to-break configuration. Figure 3d shows a photographic depiction of the tensile test experimental set-up. Six samples were tested in the directions parallel to the Langer's lines and five samples were tested in the directions perpendicular to the Langer's lines.

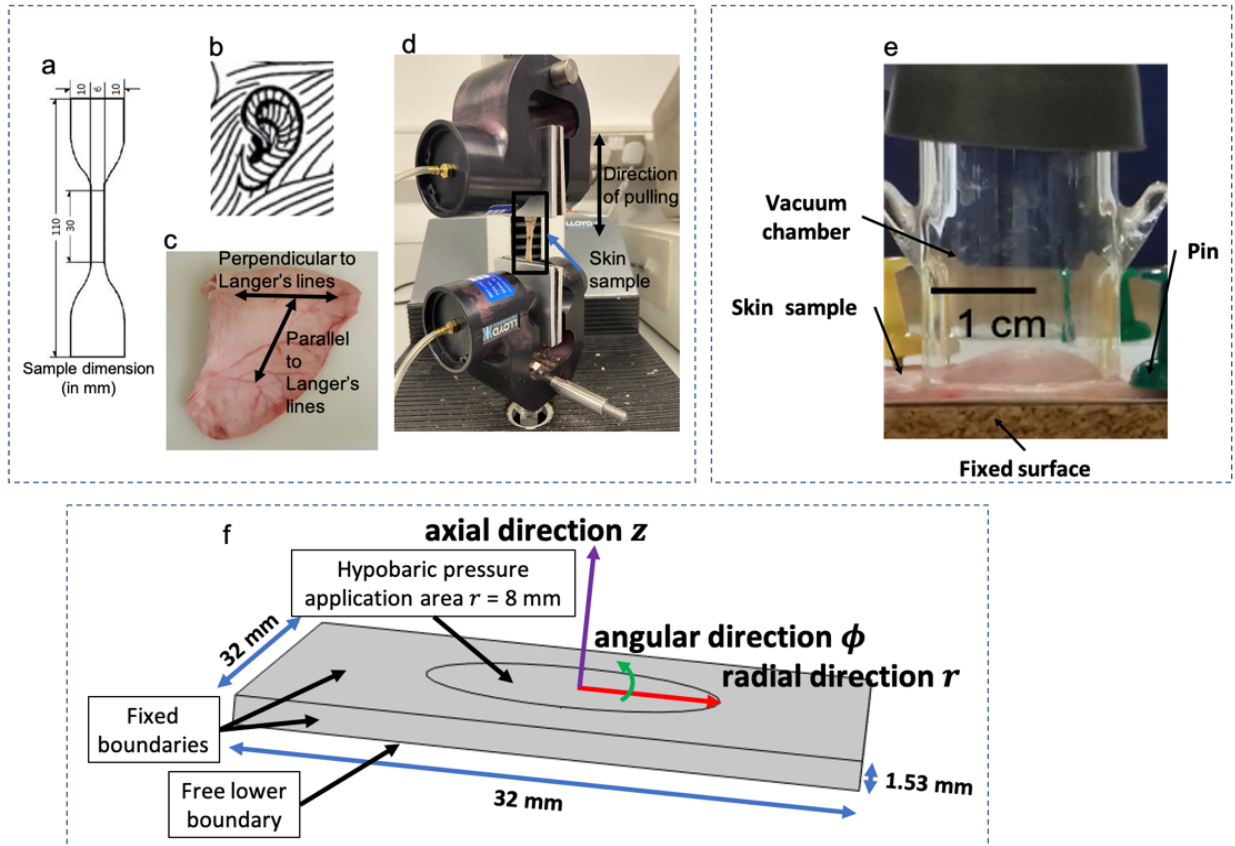

*Figure 3 Test specimen according to ASTM D-412 (a). Human ear Langer's lines (b). Directions of stretching for porcine ear skin: Parallel and perpendicular to Langer's line (c). Photograph of the tensile test apparatus (d). Photograph of the suction test experiment (e). Schematic illustration of the FEM numerical domain with its dimensions and the cylindrical coordinate system used (f).*

The stress-strain relationship obtained from the experimental tensile tests were fitted using a linear elastic model and a one-term Ogden model to account for hyperelastic theory (Ogden, 1973). Being the family of the Ogden models one of the most popular hyperelastic options for the characterisation of biological tissue (Łagan and Liber-Kneć, 2016), a one-term Ogden model provides a simple solution to fit the skin's tensile data up to the third deformation stage. For the tensile test, one can define $a_1 = a$ as the first principal stretch ratio and $\sigma_1 = \sigma$ as the principal Cauchy stress (both $a$ and $\sigma$ in the direction of elongation). The other two principal stresses $\sigma_2$ and $\sigma_3$ are equal to zero, whereas the other two principal stretch ratios can be related to $a$ as $a_2 = a_3 = a^{-\frac{1}{2}}$ when incompressibility is assumed (Li et al., 2012). Upon these conditions, one can obtain the following stress-strain relationship for a one-term Ogden formulation as

$$\sigma = \mu \left( a^{\alpha} - a^{-\frac{1}{2}\alpha} \right) \tag{1}$$

where $\mu$ and $\alpha$ are constants.

Eqn 1 was used to fit the results of the experimental tensile tests following the least-square method to obtain the values for the constants $\mu$ and $\alpha$. The constants $\mu$ and $\alpha$ of the one-term Ogden model were subsequently introduced in the FEM suction simulations.

## 2.2. Suction tests

Ex-vivo suction tests were carried out by pinning thawed 32 mm × 32 mm pig skin samples on a flat board. The thickness of the pig skin samples was in the range of $1.53 \pm 0.35$ mm. A glass compartment with an annular cross section (with an inner diameter of 16 mm and outer diameter of 19 mm) was positioned on the skin surface to produce vacuum (@ -3.5, -4.5 -6.0, and -8.3), for 1 min, as illustrated in  Figure 3e. The dome apex height was measured as the vertical distance between the dome apex and the upper surface of the pig skin sample. To obtain the dome apex height, photographs were taken with a camera placed at 170 mm from the centre point of the skin sample. Nine suction tests were performed for each vacuum pressure using 3 different areas on three different pieces of skin.

The following models can be found in the literature to obtain the Young's modulus $E$ for skin upon suction from the dome apex height $u$ and the applied pressure $p$. These models are the geometric model (De Mesquita Siqueira, 1993), which is expressed as

$$\frac{E}{1-v} = \frac{pa}{2e\left[arc\sin\left(\frac{2au}{a^2+u^2}\right) - \frac{2au}{a^2+u^2}\right]} \tag{2}$$

where $v$ is the Poisson ratio ($v = 0.48$ (Li et al., 2012)), $e$ is the thickness of the skin sample, and $a$ is the radius of the dome; the model proposed by Khatyr et al. (Khatyr et al., 2006), which is expressed as

$$\frac{E}{1-v} = \frac{0.45pa^4}{e^4\left(\frac{u}{e} + 0.3\left(\frac{u}{e}\right)^3\right)} \tag{3}$$

and Timoskenko's method (Timoshenko and Woinowsky-Krieger, 1961), expressed as

$$\frac{u}{e} + A\left(\frac{u}{e}\right)^3 = B\frac{p}{E}\left(\frac{a}{e}\right)^4 \tag{4}$$

where $A = 0.471$ and $B = 0.171$ [16].

The experimental values of the dome apex height were substituted in the above models to determine the corresponding Young's modulus $E$.

### 2.3. FEM simulations

COMSOL Multiphysics v5.5 was used to obtain the numerical FEM results in this work. A three-dimensional geometry (a rectangular domain with dimensions 32 mm × 32 mm × 1.53 mm) was created to reproduce the skin suction experiments. The geometry of the model and the boundary conditions are depicted in Figure 3f. A fixed boundary was considered on the portion of the top surface where no hypobaric pressure is applied as no entrainment of skin was observed during the experiments. Some additional simulations were also run, however, using a boundary with free horizontal movement and zero vertical displacement instead, in order to mimic the opposite ideal case where no friction between the top surface of the skin and the vacuum chamber base exists. The *Solid Mechanics* module was used. For the linear elastic, the governing equations are the conservation of momentum

$$0 = \nabla \cdot \boldsymbol{\sigma} + \boldsymbol{f} \tag{5}$$

where the superscript $\boldsymbol{f}$ is the body force per unit volume, and $\boldsymbol{\sigma}$ is the Cauchy stress tensor, the strain-displacement equations

$$\epsilon = \frac{1}{2}[\boldsymbol{\nabla u} + (\boldsymbol{\nabla u})^T] \tag{6}$$

where $\epsilon$ is the strain and $\boldsymbol{u}$ the displacement field, and the constitutive equation

$$\boldsymbol{\sigma} = \mathrm{C} : \epsilon \tag{7}$$

where $C$ is the fourth-order stiffness tensor.

As for the hyperelastic formulation, the governing equation is

$$\boldsymbol{\sigma} = \frac{1}{J}\frac{\partial W}{\partial \boldsymbol{F}} \cdot \boldsymbol{F}^T \tag{8}$$

where $W$ is the strain energy density function (which in this work will follow the Ogden equation (Ogden, 1973)), $F$ is the deformation gradient tensor, the superscript $T$ denotes transpose, and $J = \det(F)$.

The boundary conditions are as described in Figure 3f. The hypobaric pressure is applied over the central circular area, which has the same inner radius as the glass vacuum chamber (8 mm). The thickness of the glass chamber is not considered. The model assumes no initial compression on the skin caused by the positioning of the glass chamber (Khatyr et al., 2006). The lateral boundaries are fixed to mimic the effect of the pins that attach the skin to the board. The rest of the upper boundary (excluding the central circular area) is also fixed to mimic how the glass chamber blocks its movement. The central circular area and the lower boundary can experience free lateral and vertical movement. The results were all obtained by using a *Physics Controlled* mesh with the option *Extra Fine*, which gave as a result a mesh comprised of 14,410 tetrahedra. The relative tolerance of the solver was set at $10^{-8}$. The simulations were run in static mode.

### 3. Results and discussion

3.1. Tensile tests: obtaining the tensile Young's modulus and the Ogden fitting parameters

The Ogden model fitting was compared to the tensile tests, i.e. the stress $\sigma$ against the stretch ratio ($\lambda = \epsilon + 1$) both parallel and perpendicular to the Langer's lines using the experimental data from the start of the tensile test, i.e., up to a stretch ratio $\lambda = 1.2$, because the maximum stretch ratio observed in the FEM simulations of suction tests does not exceed 1.2 (see the next section) and thus the range $\lambda \in [1, 1.2]$ was shaded green (Figure 4). The parameters of the corresponding Ogden fittings, $\mu$ and $\alpha$, and the coefficient of determination $R^2$ of the fitting are shown in Table 1 and Table 2. Additional information including maximum tensile strength, failure stretch ratio, initial Young's modulus $E_0$ and maximum Young's modulus $E_{max}$, were extracted from the tests to facilitate comparison across the two data sets, i.e. parallel to Langer's lines (Table 1) and perpendicular to Langer's lines (Table 2). The standard ASTM D882–18 ("ASTM D882-18, Standard Test Method for Tensile Properties of Thin Plastic Sheeting, ASTM International, West Conshohocken, PA, 2018, www.astm.org," n.d.) recommends calculating the Young's modulus using the maximum slope of the experimental stress-strain curve, as for non-biological material the initial softer part of the curve is caused by take-up of slack, and alignment or seating of the specimen. However, for skin, the initial soft part is significant and a consequence of its own mechanical properties. Thus, Young's modulus for skin were determined from both the slope of the initial linear part and the maximum slope in the third stage, as shown in Figure 1.

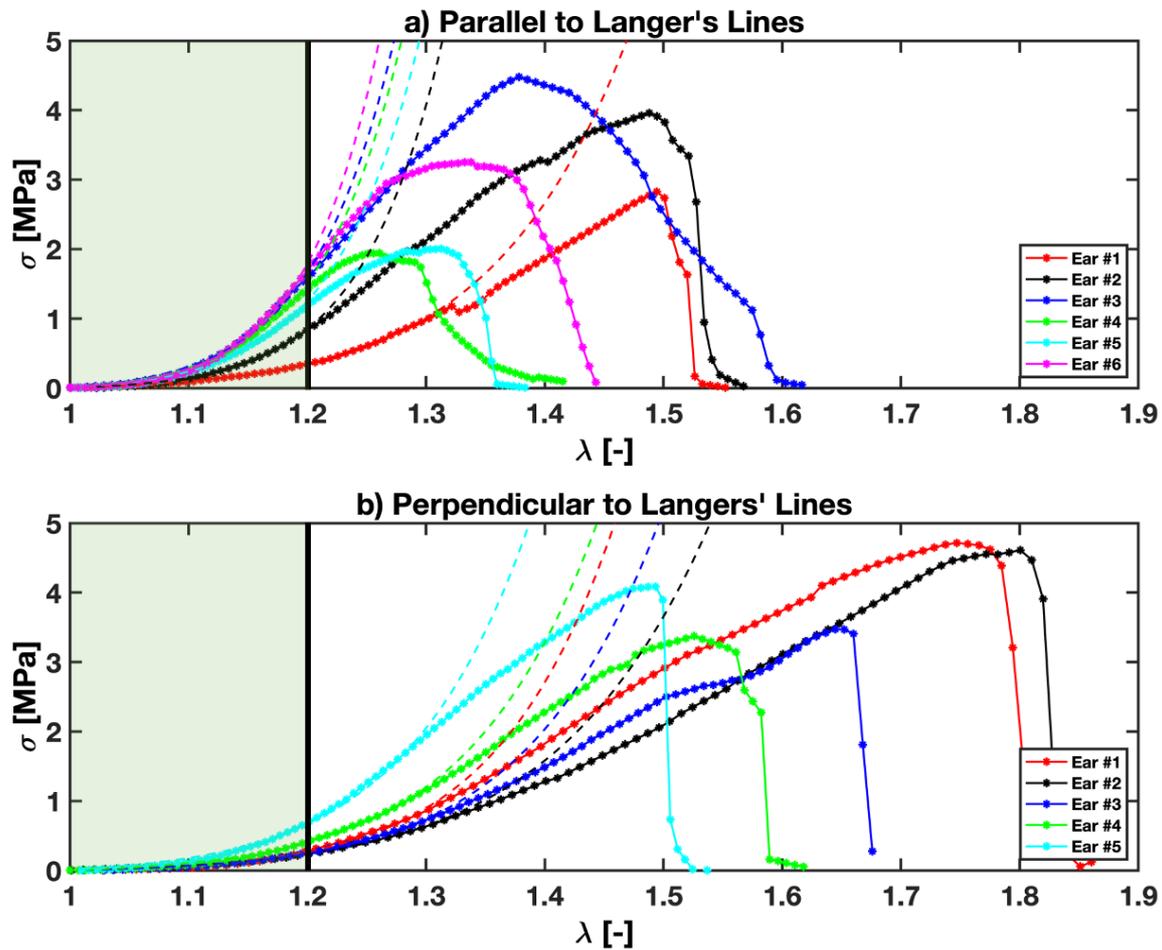

*Figure 4 Results of the experimental tensile tests and their corresponding one-term Ogden fittings for data in the range λ ∈ [1, 1.2].*

All the porcine tensile strength curves illustrated displayed the standard '3-stage' profile (Figure 4) (Ní Annaidh et al., 2012) and the maximum Young's modulus $E_{max}$, taken from the second linear portion of the profile, fall within the reported range of previous studies, which vary from 2 to 12 MPa (Gasior-Głogowska et al., 2013) and 83 MPa (Ní Annaidh et al., 2012). The results from the current study aligned most closely with Ankersen et al. (1999) (Ankersen, 1999), who reported an average Young's modulus of 14.96 MPa in porcine abdominal skin. Reported values for the initial slope in the literature are close to those obtained in this work (averages of 1.63, 1.31 and 0.98 MPa for the cases of parallel, 45° and perpendicular to the Langer's lines for human skin at a strain rate of 0.012 s⁻¹ as reported in (Annaidh et al., 2010)).

The results confirmed that the porcine ear skin is more rigid when the stress is applied in the direction parallel than perpendicular to the Langer's lines (average Young's modulus was 15.84 MPa against 12.65 MPa respectively). The influence of Langer's lines on sample orientation in tensile tests is thought to be caused by the natural orientation of the collagen and elastin fibres, which align to the Langer's lines in the dermis ((Ní Annaidh et al., 2012; Pissarenko and Meyers, 2019)). Stretching the skin parallel to the Langer's lines will therefore lead to the fibres extending further as they realign themselves in the direction of stretch (Joodaki and Panzer, 2018). The average failure stretch ratio is, however, greater for the direction perpendicular to the Langer's lines. This, combined with greater maximum stresses observed in the perpendicular than in the parallel direction, means that skin can absorb more strain energy when the stress is applied in the perpendicular than in the parallel direction to the Langer's lines. The one-term Ogden model (Ogden, 1973) fits well to both the data perpendicular to the Langer's lines and parallel to the Langer's lines, being the average correlation coefficient $R^2 \approx 0.99$ (the correlation coefficients for each separate case can be consulted in Tables 1 & 2).

*Table 1 Output tensile test variables. Direction parallel to Langer's lines*

| Variable | Ear 1 | Ear 2 | Ear 3 | Ear 4 | Ear 5 | Ear 6 | Average | SD |
|---|---|---|---|---|---|---|---|---|
| Thickness [mm] | 1.55 | 1.54 | 1.45 | 1.43 | 1.23 | 1.24 | 1.41 | 0.14 |
| Max. tensile strength [MPa] | 2.84 | 3.96 | 4.48 | 1.94 | 2.44 | 3.26 | 3.15 | 0.95 |
| Failure stretch ratio [-] | 1.50 | 1.49 | 1.38 | 1.26 | 1.31 | 1.34 | 1.38 | 0.09 |
| Maximum Young's modulus $E_{max}$ [MPa] | 16.25 | 11.34 | 19.50 | 15.07 | 12.60 | 20.32 | 15.84 | 3.60 |
| Initial Young's modulus $E_0$ [MPa] | 0.43 | 0.88 | 1.29 | 1.24 | 1.18 | 1.33 | 1.06 | 0.35 |
| $\mu$ [MPa] | 0.03 | 0.02 | 0.06 | 0.05 | 0.04 | 0.04 | 0.21 | 0.15 |
| $\alpha$ [-] | 13.08 | 19.54 | 18.59 | 18.69 | 18.27 | 20.80 | 10.38 | 2.60 |
| $R^2$ [-] | 0.99 | 0.99 | 0.99 | 0.99 | 0.99 | 0.99 | - | - |

| Variable | Ear 1 | Ear 2 | Ear 3 | Ear 4 | Ear 5 | Average | SD |
|---|---|---|---|---|---|---|---|
| Thickness [mm] | 1.36 | 1.41 | 1.53 | 1.57 | 1.49 | 1.47 | 0.09 |
| Max. tensile strength [MPa] | 4.72 | 4.62 | 3.48 | 3.38 | 4.10 | 4.06 | 0.62 |
| Failure stretch ratio [-] | 1.74 | 1.80 | 1.65 | 1.53 | 1.49 | 1.64 | 0.13 |
| Maximum Young's modulus $E_{max}$ [MPa] | 11.34 | 13.17 | 10.62 | 13.39 | 14.71 | 12.65 | 1.65 |
| Initial Young's modulus $E_0$ [MPa] | 0.54 | 0.63 | 0.66 | 0.64 | 0.92 | 0.68 | 0.14 |
| $\mu_r$ [MPa] | 0.02 | 0.03 | 0.02 | 0.03 | 0.06 | 0.03 | 0.02 |
| $\alpha_r$ [-] | 14.89 | 12.11 | 13.57 | 13.57 | 13.71 | 13.57 | 0.99 |
| $R^2$ [-] | 0.99 | 0.99 | 0.99 | 0.99 | 0.99 | - | - |

### 3.2. Comparison between experimental suction tests and FEM simulations

The Young's modulus $E$ obtained from the geometric, Khatyr and Timoshenko suction models (Table 3) using the hypobaric pressures and dome apex height obtained in the suction experiments are considerably smaller than the maximum Young's moduli $E_{max}$ from the tensile tests and of similar magnitude as the initial Young's moduli $E_0$ of the stress-strain tensile tests (Tables 1 and 2). These Young's moduli also agree with the values reported in [17].

| Hypobaric pressure [psi] | *E* [MPa]<br>(Geometric model) | *E* [MPa]<br>(Khatyr et al.) | *E* [MPa]<br>(Timoshenko) |
|---|---|---|---|
| -3.5 | 0.31–0. 90 | 1.22–2.76 | 0.34–0.84 |
| -4.5 | 0.29–1.04 | 1.17–3.29 | 0.32–0.99 |
| -6.0 | 0.30–0.73 | 1.18–2.69 | 0.32–0.77 |
| -8.3 | 0.34–0.75 | 1.33–2.89 | 0.35–0.81 |

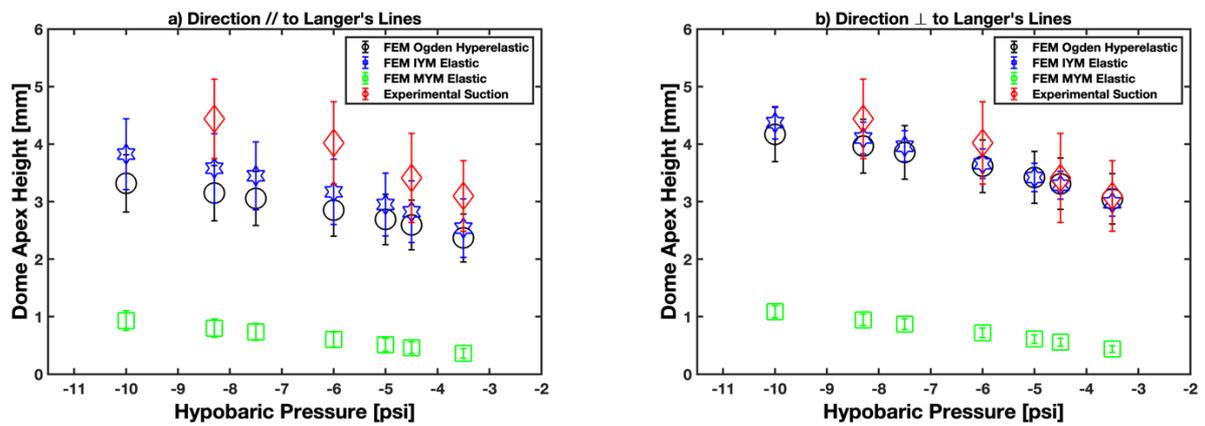

*Figure 5 Comparison between the height of the dome apex against applied hypobaric pressure obtained experimentally and FEM simulations: direction parallel (a) and perpendicular to Langer's lines (b). The FEM simulation data were obtained by using the Ogden hyperelastic model and the elastic model with the maximum Young's modulus (MYM) and the initial Young's modulus (IYM). Markers represent average while bars represent standard deviation. The experimental measurements in both graphs are the same set of data.*

FEM simulations of the suction experiments were run using three sets of parameters: (1) the linear elastic model with maximum Young's modulus $E_{max}$ from the maximum slope of the stress-strain curve of tensile tests according to ASTM standards (denoted in the graph as MYM), (2) the linear elastic model with initial Young's modulus $E_0$ (denoted on the graph as IYM), and (3) the Ogden non-linear hyperelastic model with the parameters $\mu$ and $\alpha$ fitted from the tensile tests as listed in Table 1 and Table 2. Skin is modelled as an isotropic homogeneous material with the same Young's moduli in all orthogonal directions, thus, the moduli parallel to the Langer's lines and perpendicular to the Langer's lines were considered in two separate FEM simulations for comparison.

The FEM predictions of the dome apex height are included in Figure 5 for the four pressure values tested experimentally and three additional pressure values to cover the entire range used in medical hypobaric devices. These results correspond to a fixed boundary condition. The effect of entrainment was measured for three perpendicular cases (-6.0 psi, -4.5 psi and -3.5 psi). The cases with entrainment (no friction) resulted in increased dome apex heights of 2.90%, 2.84% and 2.75% respectively. Firstly, the dome apex heights with Young's moduli parallel to the Langer's lines are slightly lower than those with Young's moduli perpendicular to the Langer's lines because of the slightly higher Young's moduli parallel to the Langer's lines due to the collagen and elastin fibres alignment in the skin. Using either a linear elastic formulation with the initial Young's modulus or the one-term hyperelastic model results in reasonable predictions of the dome apex height. Conversely, using a linear elastic formulation with the maximum Young's modulus $E_{max}$ as defined in the ASTM standards results in severe underprediction of the dome apex because using the maximum slope of the stress-strain curve as the ASTM standards specify results in the large stretch of the skin at low stress level (initial ~0.15 strain) being neglected. To this end, in the following analysis, only the linear elastic formulation with the initial Young's modulus and the Ogden hyperelastic model were considered.

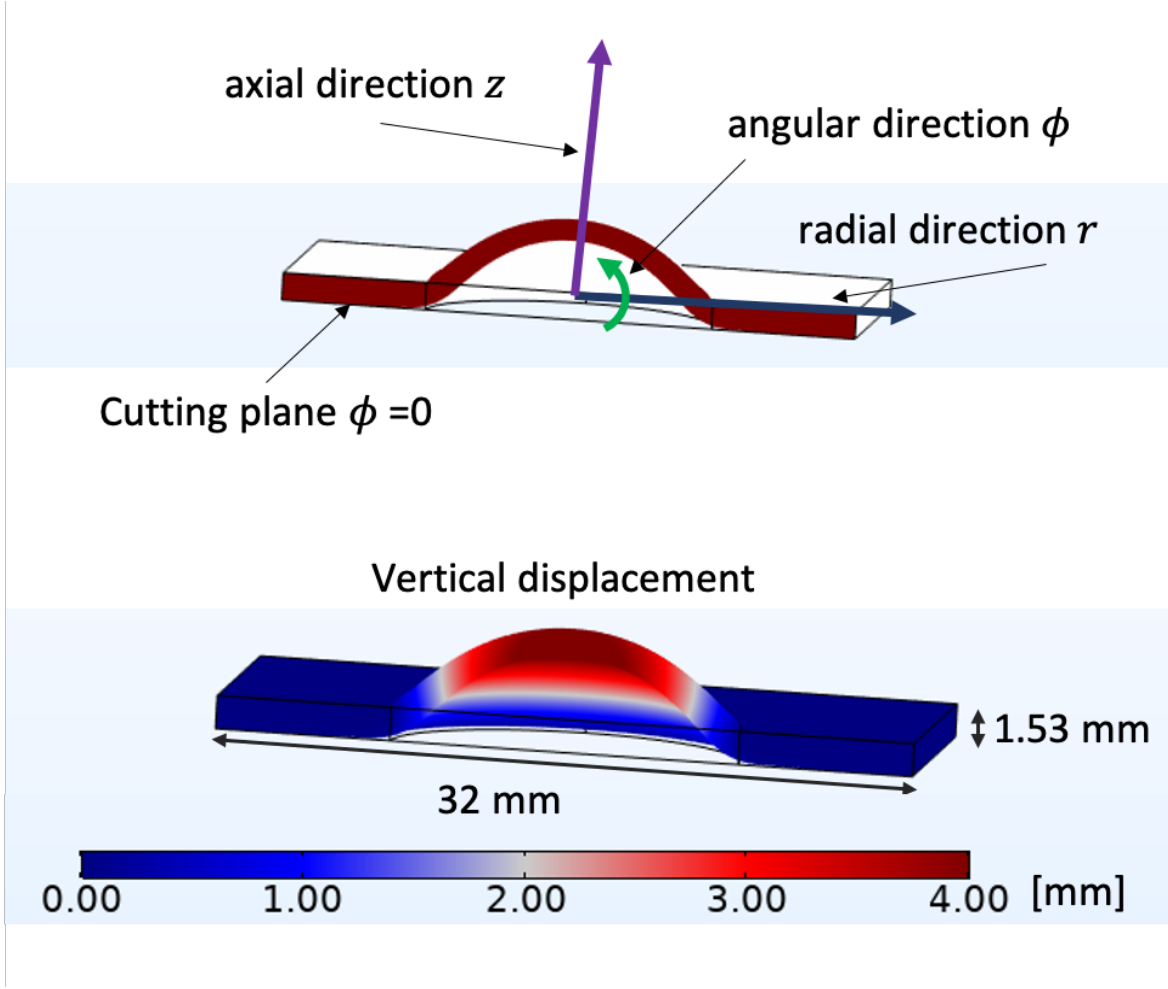

*Figure 6 The top plot shows a section of the computational domain. The cutting plane (φ=0) is highlighted in dark red. The bottom plot shows a contour map of elevation of the dome. Case with hypobaric pressure -10.0 psi. Ogden hyperelastic model (μ =0.03 MPa and α=13.57).*

Figure 6 shows an FEM visualisation of the dome formation with a contour plot of the vertical displacement. The simulated case corresponds to a non-linear hyperelastic formulation (Ogden model) with $\mu$ = 0.03 MPa, $\alpha$ = 13.57 (the average of the five fittings of perpendicular experimental cases shown in Table 2), and an applied hypobaric pressure of -10.0 psi.

### 3.3. Strain field in FEM simulations

The strain tensor $\bar{\bar{\epsilon}}$ across the suction dome in cylindrical coordinates reads

$$\bar{\bar{\epsilon}} = \begin{bmatrix} \epsilon_{rr} & 0 & 0 \\ 0 & \epsilon_{\phi\phi} & 0 \\ 0 & 0 & \epsilon_{zz} \end{bmatrix}, \tag{9}$$

where $\epsilon$ is the strain and the subindices $r$, $\phi$, and $z$ denote the cylindrical coordinates as depicted in Figure 3f and Figure 6. The shear strain components were negligible (i.e. $\epsilon_{r\phi} = \epsilon_{rz} = \epsilon_{\phi z} \cong 0$) as confirmed by the simulation output. The non-zero components of the strain tensor are shown as a function of the distance from the centre of the dome (vacuum application area) in Figure 7 for the case with an initial Young's modulus $E_0 = 0.68$ MPa (an average of $E_0$ for the five cases in the direction perpendicular to Langer's lines) and in Figure 8 for the non-linear Ogden formulation with $\mu = 0.03$ MPa and $\alpha = 13.57$ (the average of the five fittings perpendicular to Langer's lines) in order to highlight the differences between them. A linear elastic model with $E = 0.68$ MPa gives similar dome apex height as using the non-linear Ogden model with $\mu = 0.03$ MPa and $\alpha = 13.57$ (i.e. 3.19 mm, 3.72 mm and 4.12 mm for the linear elastic method against 3.41 mm 3.85 mm and 4.16 mm for the Ogden model for the three hypobaric pressures represented). The results in both figures are taken on the cutting plane shown in Figure 6, that is, along the radial axis $r$ of the cylindrical system, angular coordinate $\phi=0$, and initial vertical coordinate $z=0$ (bottom surface of the skin) and $z = 1.53$ mm (top surface of the skin). By representing the strain components along the upper and lower surface, one can determine whether there is a stretching (positive strain), compression (negative strain), or a transition between them (i.e. a neutral line) in any of the directions. The vertical black line denotes the boundary between the vacuum application area and the rest of the skin sample, i.e., the blue shaded area denotes the area where the hypobaric pressure is applied.

A certain degree of distortion and stress concentration are observed along the boundary where there is an abrupt change from the pressurised area to the non-pressurised area for both the linear elastic and the non-linear Ogden formulation. The stress concentration is more severe along the dome rim with higher hypobaric pressure, in particular for linear elastic model. High stress concentration or skin distortion can lead to damage to skin, which should be restrained carefully. One can also see that the stretch ratio in the axial coordinate $\epsilon_{zz}$ is smaller than 1.2 across the dome, which validates the choice for the range of stretch ratios used to obtain the Ogden fittings in Figure 4. Stretch ratio in most dome top surface is around 1.2 (=1+$\epsilon_{zz}$), which shows the skin top surface has passed the initial softer stages and entered the transition or stiffened stage of the stress-strain curve.

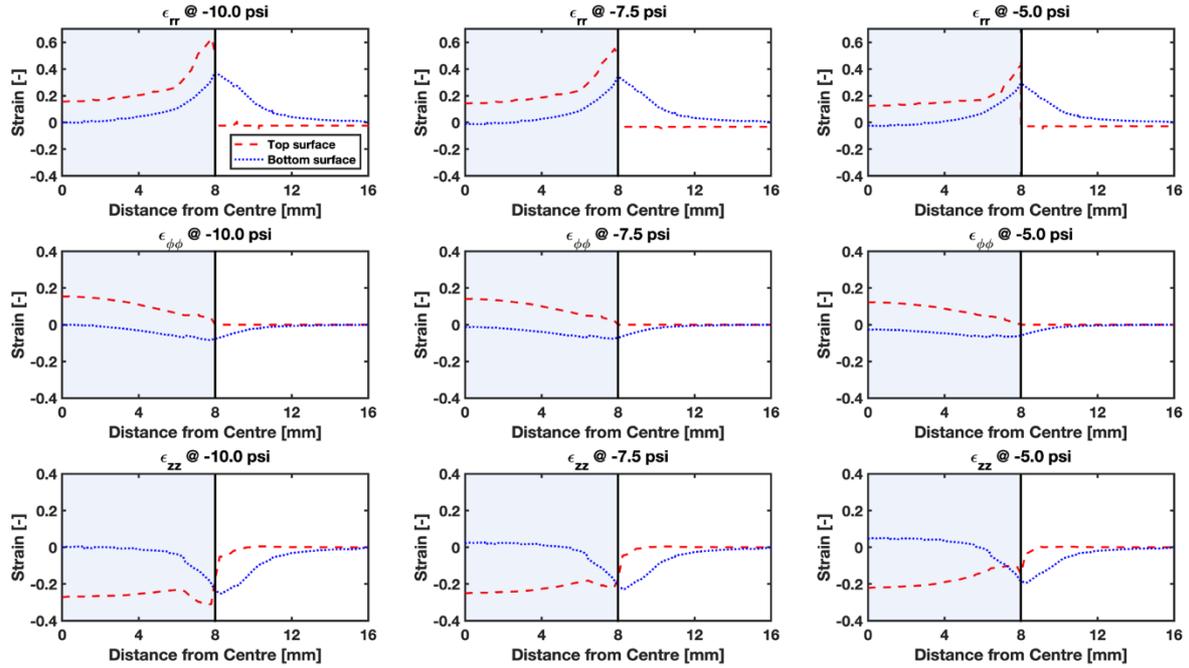

*Figure 7 Cylindrical components of the strain tensor on the upper and lower surface of the skin during suction test. Results for FEM linear elastic model E = 0.68 MPa. The upper limit of the y-axis for the radial strain component $\varepsilon_{rr}$ has been set at 0.7 to better visualise the reach of the effect of concentration of stresses in the area.*

It can be observed from Figures 7&8 that the top surface of the skin is mainly stretched in the radial (~20%, in transition to stiffened stage in tensile stress-strain curve) and angular (~10%, in transition stage) directions, while the bottom of the skin only develops a stretch (within softer stage) near the edge of the vacuum area. It can be inferred that along the thickness of the skin the radial stretch ratio is reduced from the top to the bottom of the skin, which can be confirmed by Figure 9, where a section plot of the 3D visualisation of the principal strain distributions is shown. The circumferential/angular stretch ratio shows similar trend, but beyond $r = 4$ mm a slight compression develops along the bottom. It is notable that doubling vacuum pressure from -5psi to -10psi only increases the strain in the skin slightly.

As observed from Figure 9, most region along the thickness, except for the region close the bottom surface, was stretched in the radial direction by at least 10%, which is beyond the initial linear elastic stage of the stress-strain relationship. In the $\phi$-direction, the crown of the dome, except the bottom surface, is stretched by at least 10%, while along the rim of the dome, the inner/bottom half of the skin is in compression and the outer/top half is in tension. In the z direction (along the thickness), skin is in compression with decreasing strain magnitude from the top (~25%) to the bottom surface (~0%), indicating that the thickness of the skin is reduced slightly under vacuum.

Figure 10 confirms this trend along the thickness by showing the vertical displacement of the top and bottom surface of the skin. Greater upward displacements are observed for the bottom surface, which implies that contractions in thickness of 0.2mm ~0.3 mm, that is 13%~20% of the initial thickness (1.53 mm), occur at the centre of the domes. This is in accordance to the strains observed in the axial direction $\epsilon_{zz}$ in Figure 8.

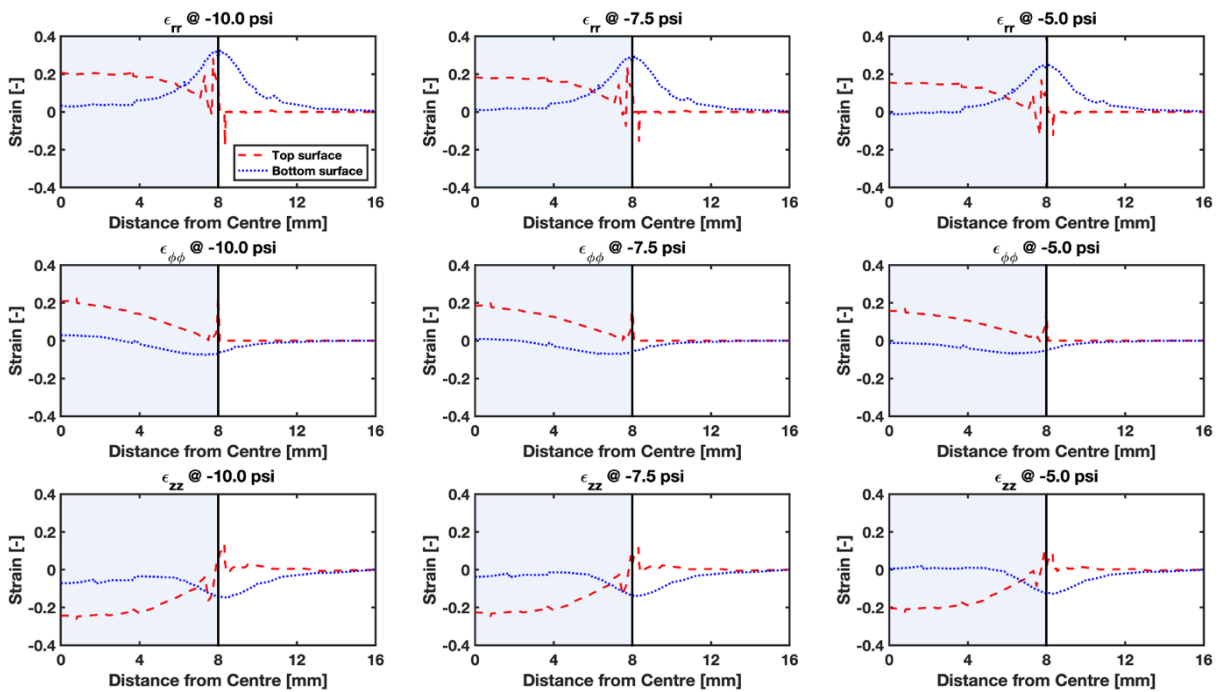

*Figure 8 Cylindrical components of the strain tensor on the upper and lower surface of the skin during suction test. Results for FEM Ogden non-linear elastic model μ = 0.03 MPa and α = 13.57 (average values for the Ogden fittings to the tensile tests perpendicular to Langer's lines).*

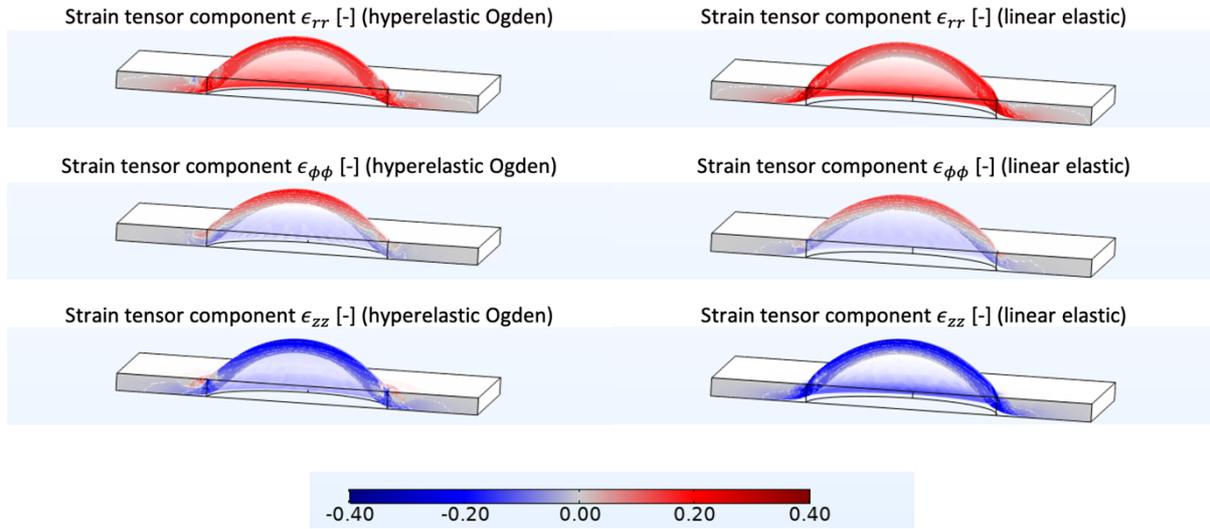

*Figure 9 A section view showing the distribution of the principal strain components (section plane contains the origin of coordinates and is parallel to the radial-axial plane). Results obtained using FEM (Ogden non-linear elastic model $\mu = 0.03$ MPa and $\alpha = 13.57$) at -5 psi on the left-hand side column. Results obtained using the linear elastic model (E=0.68 MPa) on the right-hand side.*

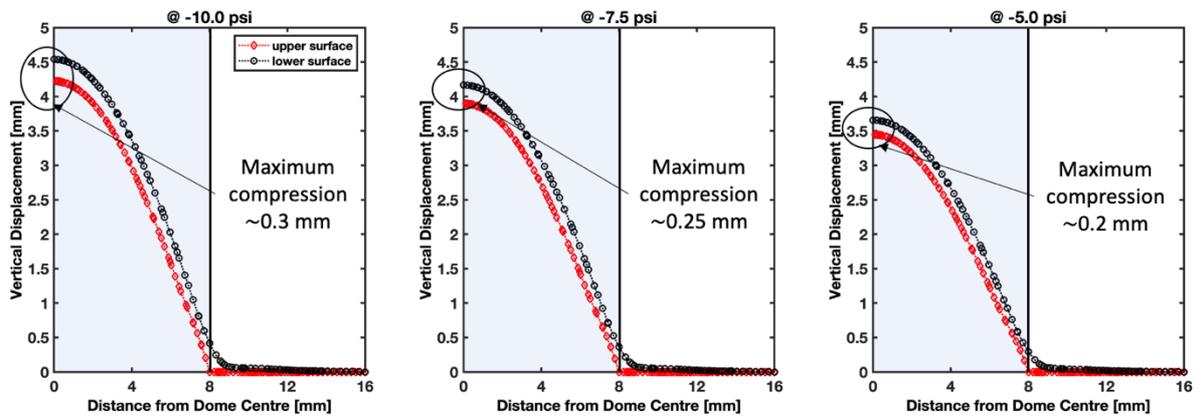

*Figure 10 Plot of vertical displacement for the upper and lower surface of the skin. Non-linear elastic (Ogden) case ($\mu=0.03$ MPa $\alpha=13.08$).*

Also, by comparing Figure 7 and Figure 8 as well as Figure 9, it is found that the trends vary depending on whether a linear elastic or a non-linear Ogden formulation is used. This is evident for the radial strain $\epsilon_{rr}$, which increases greatly nearby the boundary when a linear elastic formulation is used because concentration of stresses occurs in that area. However, with a non-linear elastic (Ogden) formulation, increase of stresses leads to increase of stiffness and the consequent strain actually decreases towards the boundary, giving rise to a smoother transition between the area of applied pressure and the rest of the simulated skin. The stretch ratio of the skin is of paramount importance for assessing the potential skin damage during application of hypobaric-assisted medical devices, thus the more accurate hyperelastic model is desired for designing/tuning safe hypobaric pressure applied on skin via medical devices.

## 4. Conclusions

A review of the available literature reveals different ranges of the Young's modulus for skin depending on whether suction or tensile deformation tests are applied. The apparent mismatch between suction and tensile Young's modulus can be explained by the magnitude of the strain developed upon suction deformation mode, which falls within different stages of the theoretical tensile stress-strain curve at different locations across the dome.

In the present study, a combination of FEM simulation and experimental suction and tensile tests illustrates that using the Young's modulus derived from the initial slope of the stress-strain tensile curve on pig skin gives reasonable predictions of the dome apex height. Simultaneously, fitting the stress-strain curve to a one-term Ogden hyperelastic model results in good predictions of the dome apex height.

Although useful for the prediction of the dome apex height, the limitations in the use of the initial Young's modulus to analyse the mechanics of the dome formation become evident when carrying out an analysis of the strain distribution across the suction dome as skin presents a non-linear rather than linear strain-strain characteristic. Comparing the strain distributions with the two methods (either linear elastic or hyperelastic formulation) shows different trends on the calculated strain across the dome, particularly on the radial component.

The analysis of the strain distribution over the suction dome using the hyperelastic formulation reveals that:

- At the centre of the dome, the radial strain decreases from the top surface to the bottom surface, although always in tension across the dome. No compression is observed on the radial direction across the suction dome.

- Analogous behaviour is observed for the angular strain component on the top side of the skin. The bottom side develops tension in the crown area up to approximately $r = 4$ mm, beyond which compression appears.

- Compression is also observed along the thickness direction, which decreases from the top to the bottom of the skin. Consequently, the thickness of the skin decreases upon suction.

The observations of strain distributions are valuable for guiding the future design of hypobaric-assisted drug delivery devices and controlling the potential skin damage caused by these devices.

**Acknowledgements**

The authors would like to acknowledge the UK Engineering and Physical Sciences Research Council (EPSRC) for the financial support (grant number: EP/S021167/1 and EP/S021159/1).